\documentclass[preprint]{aastex}

\usepackage{natbib}
\slugcomment{\textbf{To be published in the Astrophysical Journal}}

\shorttitle{Mid-IR Spectra of SN 2014J}

\shortauthors{Telesco et al.}

\begin{document}

\title{MID-IR SPECTRA OF TYPE IA SN 2014J IN M82 SPANNING THE FIRST FOUR MONTHS}

\author{Charles M. Telesco\altaffilmark{1}, Peter H\"{o}flich\altaffilmark{2}, Dan Li\altaffilmark{1},
Carlos \'{A}lvarez\altaffilmark{3,4}, Christopher M. Wright\altaffilmark{5}, Peter J. Barnes\altaffilmark{1}, 
Sergio Fern\'{a}ndez\altaffilmark{3}, James H. Hough\altaffilmark{6}, N. A. Levenson\altaffilmark{7}, Naib\'{i} Mari\~{n}as\altaffilmark{1}, 
Christopher Packham\altaffilmark{8}, 
Eric Pantin\altaffilmark{9}, Rafael Rebolo\altaffilmark{3}, Patrick Roche\altaffilmark{10}, Han Zhang\altaffilmark{1}}
\altaffiltext{1}{University of Florida, Department of Astronomy, Gainesville, Florida, 32611, USA, telesco@astro.ufl.edu}
\altaffiltext{2}{Florida State University, Department of Physics, Tallahassee, Florida, 32305, USA, phoeflich77@gmail.com}
\altaffiltext{3}{Instituto de Astrof\'{i}sica de Canarias, C/ V\'{i}a L\'{a}ctea, s/n E38205 La Laguna (Tenerife), Spain}
\altaffiltext{4}{Universidad de La Laguna, Pabell\'{o}n de Gobierno, C/ Molinos de Agua s/n 38200 La Laguna, Spain}
\altaffiltext{5}{School of Physical, Environmental and Mathematical Sciences, University of New South Wales, PO Box 7916, Canberra BC ACT 2610, Australia}
\altaffiltext{6}{School of Physics, Astronomy and Mathematics, University of Hertfordshire, Hatfield, Hertfordshire, AL10 9AB, UK}
\altaffiltext{7}{Gemini Observatory, Casilla 603, c/o AURA, La Serena, Chile}
\altaffiltext{8}{Physics \& Astronomy Department, University of Texas at San Antonio, 1 UTSA Circle, San Antonio, Texas 78249}
\altaffiltext{9}{Service d'Astrophysique CEA Saclay, France}
\altaffiltext{10}{Department of Astrophysics, Oxford University, Denys Wilkinson Building, Keble Road, Oxford, OX1 3RH, UK}

\begin{abstract}
We present a time series of 8 - 13 $\micron$ spectra and photometry for SN 2014J obtained 57, 81, 108, and 137 d after the explosion using CanariCam on the Gran Telescopio Canarias. This is the first mid-IR time series ever obtained for a Type Ia supernova. These observations can be understood within the framework of the delayed detonation model and the production of $\sim$0.6 $\rm M_\odot$ of $^{56}$Ni, consistent with the observed brightness, the brightness decline relation, and the $\gamma$-ray fluxes. The [Co III] line at 11.888 $\micron$ is particularly useful for evaluating the time evolution of the photosphere and measuring the amount of $^{56}$Ni and thus the mass of the ejecta. Late-time line profiles of SN 2014J are rather symmetric and not shifted in the rest frame. We see Argon emission, which provides a unique probe of mixing in the transition layer between incomplete burning and nuclear statistical equilibrium. We may see [Fe III] and [Ni IV] emission, both of which are observed to be substantially stronger than indicated by our models. If the latter identification is correct, then we are likely observing stable Ni, which might imply central mixing. In addition, electron capture, also required for stable Ni, requires densities larger than $\sim$$1 \times 10^9$ g cm$^{-3}$, which are expected to be present only in white dwarfs close to the Chandrasekhar limit. This study demonstrates that mid-IR studies of Type Ia supernovae are feasible from the ground and provide unique information, but it also indicates the need for better atomic data.
\end{abstract}


\section{INTRODUCTION}\label{Introduction}
Type Ia supernovae (SNe~Ia) have proven invaluable in cosmological studies, understanding the origin of the elements, and as laboratories for probing the physics of flames, instabilities, radiation transport, non-equilibrium systems, and nuclear and high energy physics. There is general agreement that SNe~Ia are the thermonuclear explosions of C/O white dwarfs (WDs) and that the competition between nuclear burning and hydrodynamical time scales determines the result of the explosion. To first order, the outcome hardly depends on details of the physics, the scenario, or the progenitor evolution (``stellar amnesia''), because nuclear physics governs the structure of the progenitor and the explosion, and radioactive $^{56}$Ni powers the light curves, which we subsequently observe as SNe~Ia \citep{HGL03}. As a result, the apparent homogeneity of SNe~Ia does not imply homogeneous classes of either progenitors or explosion scenarios. However, advances in observations and computational methods are allowing detailed studies of secondary effects \citep[e.g.,][]{branch99}, which can be crucial for distinguishing among alternatives. For example, with time, the expanding envelope becomes increasingly transparent and, thus, spectral time series can probe the structure of the envelope and the corresponding radial structure of the exploding object. The inferred distribution of products of explosive C burning (O/Mg/Ne), incomplete burning (Si/S and Ca/Ar), and complete burning then provides a powerful link to the theory. Timely observations of nearby supernovae provide an outstanding opportunity to advance the field. 

Supernova 2014J was discovered on January 21.81 UT by \citet{Fossey14}  using a 0.35-m telescope while imaging M82 in the B, V, and R bands. The inferred date of the explosion is January $14.72 \pm0.2$ \citep{Zheng14}, and it reached maximum B-band brightness $\sim$18 days later, on February $1.74 \pm 0.13$ \citep{Marion14}. SN~2014J appears to be a relatively normal SN~Ia, although it exhibits somewhat high extinction ($\rm A_V \approx 2$ mag, \citealt{foley13,Marion14}), and some higher-than-expected velocity features \citep{Goobar14_94J}. The distance of the host galaxy M82, $3.53 \pm 0.26$ Mpc \citep{Karachentsev06}, makes SN~2014J the closest SNe~Ia in decades, and it has been observable continuously over many months starting only a few hours after the explosion \citep{Marion14}. The result has been a wealth of data including high-quality spectra and light curves from the optical to the near-IR, tracking of the detailed UV evolution \citep{foley14}, tight limits on the X-ray fluxes, and, for the first time on any SN Ia, detection of $\gamma$-rays \citep{Isern14,Churazov14}.

In this paper, we present the time-dependent evolution of the mid-IR spectrum of SN~2014J, in fact the first ever for any SN~Ia. Prior to SN~2014J, late-time mid-IR spectra had been obtained only for SN~2005df and SN~2003hv at $\sim$135 d and  $\sim$375 d, respectively, using the \it Spitzer Space Telescope \rm \citep{gerardy07}. Given this special opportunity and the important role that SNe~Ia play in our understanding of both stellar evolution and cosmology, we initiated a campaign of photometry and spectroscopy of SN~2014J in the 8 - 13 $\micron$ region using CanariCam at the 10.4-m Gran Telescopio Canarias (GTC). Opening a new wavelength range presents both an opportunity and a challenge to identify features and understand their evolution. The mid-IR spectral region offers the practical advantage of being substantially less affected by interstellar extinction than the optical and near-IR, and, as we show, it offers unique access to supernova properties fundamental to understanding SN~2014J and its class \citep{gerardy07}. \citet{li2014sn2014j_2, li2014sn2014j_1} presented early results of this program. 

\section{OBSERVATIONS \& DATA}\label{Observations}
\noindent \textit{\textbf{Procedures:}}
CanariCam is the GTC facility mid-IR, multi-mode camera developed by the University of Florida (\citealt{Telesco03}; http://www.gtc.iac.es/instruments/canaricam/canaricam.php). It employs a $320\times240$-pixel Raytheon detector array with 0\farcs08 pixels, which provides a field of view of 26\arcsec\,$\times$\,19\arcsec\,with Nyquist sampling (two pixels per $\lambda/D$) of the diffraction-limited point-spread function at 8 $\micron$. For the observations presented here, we used the standard imaging mode for photometry and the 10-$\micron$ low-spectral-resolution spectroscopy mode (LoRes-10) with a 1\farcs04-wide slit positioned along the N-S direction. For the imaging/photometry we used the Si2 ($\lambda = 8.7$ $\micron$, $\rm \Delta \it \lambda \approx$1.1 $\micron$), Si4 ($\lambda = 10.3$ $\micron$, $\rm \Delta \it \lambda \approx$0.9 $\micron$), and SiC ($\lambda = 11.75$ $\micron$, $\rm \Delta \it \lambda \approx $2.5 $\micron$) filters. We used standard chop-nod procedures with an 8\arcsec\,chop-nod throw. A log of observations is presented in Table 1.

The observations were processed in \it iDealCam\rm, a custom IDL package developed to reduce CanariCam data \citep{Li13}. All chop frames in raw FITS files were checked by eye, and those few with apparent artifacts (high-background residuals, poor seeing, etc.) were discarded. We used a fixed aperture of 19-pixel (=1\farcs52) width to extract the spectra. Flux calibration and telluric correction were achieved using the spectral template of a nearby Cohen standard HD 92523 \citep{cohen1999}. Wavelength calibration was done using several sky lines identified in the raw images. Deliberately shifting the spectrum by as much as $\pm$2 pixels (0.04 $\micron$) from the best-fit solution resulted in obvious mismatches between the sky lines for the Cohen standard star and those in the raw SN 2014J frames; we therefore consider the uncertainty in wavelength calibration to be $\pm$0.04 $\micron$. Using the sky lines, we estimated the spectral resolution to be $\sim$0.16 $\micron$, corresponding to 4800 km s$^{-1}$.

Our spectra span most of the 10 $\micron$ atmospheric window. Increasing noise at each end of the spectra results from a combination of decreasing transmission and increasing thermal emission and sky noise at the window edges. The spectral blocking filter enhances the former effect by cutting off fairly sharply below 7.8 $\micron$ and above 12.9 $\micron$. In addition, strong atmospheric ozone absorption occurs in the range 9.3 - 10.1 $\micron$, with correspondingly higher noise there. The net result is that useful data span the approximate range 7.8 - 12.9 $\micron$, with diminished quality near the ends of this range and in the vicinity of the ozone feature.

We began our mid-IR campaign on 2014 February 10 and 20 with exploratory mid-IR imaging, which revealed the supernova to be at the nominal position \citep{Fossey14} and bright enough that mid-IR spectroscopy would be feasible with CanariCam. The flux ratios for the narrow-band filters also indicated that a bright line might be present in the longer-wavelength filter, which further motivated our spectroscopy. Even though M82 has a highly structured mid-IR-bright starburst spanning the central 300 pc \citep{Telesco92,Gandhi11}, SN~2014J is well outside of the main starburst region and, with the $\sim$0\farcs3 resolution of CanariCam at the GTC, was identified unambiguously; no other point sources were detected in the CanariCam field of view. We were able to obtain spectra on four separate occasions at intervals of roughly one month during the period 57 to 137 d after the explosion. On all but the last of those occasions we also obtained photometric images to provide a basis for extrapolating the spectra, at least coarsely, to the two dates prior to our use of the spectroscopic mode. The reduced spectra, both smoothed and unsmoothed, are shown in Figure 1.

\noindent \textit{\textbf{General Properties:}}
We obtained mid-IR spectra of SN~2014J at 57, 81, 108 and 137 d (Figure 1). We indicate the rest wavelengths of several forbidden emission lines expected to be prominent based on detailed modeling presented in \S \ref{Models} and on \it Spitzer \rm spectroscopy of SN~2005df at 135 d by \citet{gerardy07}. The spectra are dominated by Co and weaker lines of Fe, Ar, and Ni primarily in the second and third ionization stages. The strongest feature by far in all of the spectra is the [Co III] line with a rest wavelength of 11.888 $\micron$. While there is some low-level blending in the wings, the feature is relatively isolated and easily characterized, which constitutes a real benefit of this spectral region. The half-width of the line across the range of observation dates is about 6800 km s$^{-1}$, being slightly narrower (6400 km s$^{-1}$) on day 137. On all dates, line wings extend out to $\sim$10,000 km s$^{-1}$. 

After the [Co III] line, the next most prominent features are in the vicinity of the [Co II] 10.523 $\micron$ line, which is likely blended with the [S IV] 10.510 $\micron$ line. Unique to the mid-IR region is the [Ar III] line expected at 8.991 $\micron$. Ar is a primary tracer for transition layers between incomplete O burning and nuclear statistical equilibrium (NSE, see \S \ref{Models}). Ar is expected to be exposed early, with little change in flux thereafter, just as we observe. While not well isolated, its width appears to be comparable to that of the [Co III] line. Finally, we identify a broad bump in the 8.0-8.4 $\micron$ region as probably a blend of the [Fe III] 8.211 and 8.611 $\micron$ lines and, starting on day 81, the [Ni IV] 8.405 $\micron$ line. 

Prior to our observations, the only mid-IR spectra of SNe Ia were those of SN 2003hv and 2005df obtained with \it Spitzer \rm at 375 d and 135 d, respectively, after the explosions \citep{gerardy07}. Considering the fortuitous coincidence of observation dates, further comparison of our spectrum of SN 2014J on day 137 with that of SN 2005df is worthwhile. In Figure 2 we show both spectra, with the peak [Co III] 11.888 $\micron$ flux density of SN 2005df scaled to that of SN 2014J. The resolutions of the two spectra are comparable: $\sim$0.15 and $\sim$0.10 $\micron$ for SN 2014J and 2005df, respectively, which are 3-5 times narrower than the FWHM of the [Co III] line. Thus, while the spectra are slightly smoothed, both have features that are well resolved and can be compared without regard to the difference in instrumental resolution. The spectra, taken at almost the same time in the evolution of the light curves, display many of the same features. Most striking perhaps are the almost identical profiles displayed by the [Co III] 11.888 $\micron$ lines, which have approximate half-widths of 6400 km s$^{-1}$. Also notable, the strengths of the [Ni IV] and [Ar III] lines relative to the [Co III] line are comparable in the two spectra, except for the ``two-pronged'' [Ar III] feature seen in SN 2005df. \citet{gerardy07} attribute that feature to a non-spherically symmetric emission component. Alternatively, it could be due to self absorption in the line. We note incidentally that the [Co III]-line scaling factor of 33 (Figure 2) corresponds to a distance to SN2005df of 20.1 Mpc if the line luminosities are the same. This value is on the slightly high end, but nevertheless within the range, of distances derived by various other methods for the host galaxy NGC 1569 \citep{gerardy07,brown2010,milne2010}; this may suggest an interesting secondary use for these lines.

We are reasonably confident of the above identifications in SN 2014J. However, there are hints of other, very weak and blended lines that are marginally detected at best, being only apparent in the smoothed versions of the spectra. These include: (1) what may be the [Co II] 11.167 $\micron$ line, seen most strongly on day 81, but seeming to persist on all dates at some level; (2) the [Ni II] 12.729 $\micron$ line, which may be blended with the [Co III] 12.681 $\micron$ line and may persist on all four dates but, perhaps due to noise, does not appear entirely constant or consistent in wavelength; (3) a feature at 12.4 $\micron$, which may emerge from the wing of the strong [Co III] line on day 137; and (4) a bump on day 137 near 9.7 $\micron$ associated with a ``dip'' at 9.5 $\micron$ that, we speculate, could be a P Cygni profile associated with the [Ni IV] 9.723 $\micron$ line. Of these features, the one near 11.167 $\micron$ is the most likely to be real, although we cannot be sure of the identification. The others are in noisy regions of the spectra and may not be real. We do note that Gerardy et al. (2007) may have also detected a feature in SN~2005df at a position roughly that of the [Ni II] 12.729 $\micron$ line (Figure 2). 

\noindent \textit{\textbf{Temporal Variations:}} 
Some parts of the spectrum remained virtually unchanged across our observations, in particular, the 8.0 - 9.5 $\micron$ region where the [Ni IV] and [Ar III] lines are expected. On the other hand, [Co II] at 10.523 $\micron$ stayed constant between days 57 and 81 and subsequently decreased by about 50\%. Likewise, a dramatic evolution was seen in the strength of the [Co III] feature at 11.888 $\micron$, which increased in strength by about 50\%, then remained roughly constant or decreased slightly. [Co III] is an excellent tracer of the amount of Co exposed above the photosphere and below the critical density for collisional de-excitation. 

In Figure 3a, we plot the photometry determined directly from the images taken with the CanariCam Si2 and SiC filters. In Figure 3b, we show the evolution in the line fluxes measured from the spectra. No spectra were taken for the first two dates (days 26 and 36) in Figure 3b. However, since the SiC filter spans the entire [Co III] 11.888 $\micron$ line and a large portion of the [Co II] 10.523 $\micron$ line, we are able to extrapolate the line fluxes to those earlier dates if we make an assumption about the line ratio (we assume the [S IV] line contributed negligibly). Specifically, we assume that the [Co II]/[Co III] flux ratio was somewhere between unity and the value measured on day 57 (i.e., when the first spectrum was obtained), which gives the ranges of possible [Co II] and [Co III] fluxes for days 26 and 36 indicated by the shaded areas in Figure 3b. Taken at face value and assuming that the [Co III] line flux was proportional to the exposed Co mass, we are tempted to infer that 26\% of the Co was exposed on day 26, with a nearly linear increase in exposed mass until full Co exposure around day 81. However, our models indicate that only about 3\% of all Co was below the critical density on day 26, and the flux on that day consisted mainly of continuum. Thus, the forbidden line of [Co III] on day 26 was weaker by at least a factor of four than it was on day 57. 

The [Co III] 11.888 $\micron$ line also exhibited changes in the profiles. It is evident in Figure 1 that the profile on day 57 was not merely a re-normalized version of that on day 137, but rather a ``stubby'' version with a somewhat truncated peak. In addition, compared to the profile on day 137, the centroids of the profiles on days 81 and 108 may be shifted slightly blue-ward and red-ward, respectively. These profile shifts are comparable to the uncertainty in the wavelength calibration and therefore, taken by themselves, are not significant. However, for both days the FWHM of the lines were also broader than on day 137. Thus, we may be seeing some combination of changes in both line widths and positions.  

\section{MODELS}\label{Models}
Based on detailed, spherical models for supernovae discussed below, we analyzed SN~2014J's mid-IR spectra and their evolution. Given the initial WD model and parameterized properties for nuclear burning, the time evolution follows without additional free parameters. 

\noindent \textit{\textbf{Methods:}}
For the calculations of explosions, light curves, and spectra we used our code for radiation transport \citep[HYDRA;][]{h90,h95,H02m,h09}, which solves for the hydrodynamics using the explicit Piecewise Parabolic Method \citep[PPM;][]{CW_PPM84}, detailed nuclear and atomic networks \citep{Cy10,k94d,h95,seaton2005}, transport for low-energy and $\gamma$-photons and positrons by variable Eddington Tensor solvers and Monte Carlo Methods \citep{mm84,stone1992,h93,H02m,h09,penney14}. For this study, atomic data for forbidden line transitions have been updated using \citet{soma10a}, \citet{brian_11fe14}, and \it The Atomic Line List \rm (V2.05B18, http://www.pa.uky.edu/~peter/newpage/) by Peter van Hoof.

\noindent \textit{\textbf{General Scenarios:}}
The consensus is that SNe~Ia result from a degenerate C/O WD undergoing a thermonuclear runaway \citep{hf60} and that they originate from close binary stellar systems. Potential progenitor systems may consist of either two WDs, a so-called double degenerate system (DD), and/or a single WD and a main sequence, helium, or red giant star, a so-called single degenerate system (SD). For overviews see \citet{branch95}, \citet{nomoto03}, \citet{wang2012}, \citet{Stephano11}, and \citet{Stephano12}.

Within this general picture, the explosion of the WD is triggered either by heat release during the dynamical merging of two WDs \citep{webbink84,iben,benz90} or by compressional heating when the WD approaches the Chandrasekhar limit ($\rm M_{Ch}$). Theoretical work on the explosion, spectra, and light curves seem to favor $\rm M_{Ch}$ explosions whether originating from SD or DD progenitors, with some contribution of dynamical mergers 
\citep{hk96,sainom98,sainom85,shen_mergers11}. For the $\rm M_{Ch}$ case, the most likely scenario involves delayed-detonation models \citep{khokhlov91,woosley94,yamaoka92,gko04a,p11}, i.e., so-called delayed-detonation-transition (DDT) models, which incorporate transitions from a subsonic deflagration front to a supersonic detonation front.  

\noindent \textit{\textbf{Selection of the Reference Model:}}
The sequence of optical spectra obtained by \citet{Marion14} indicates that SN2014J had a layered structure 
with little or no mixing in the outer layers, consistent with delayed detonation. Burning of the WD's C/O 
mixture would produce regions of iron-group elements in the densest regions where there is NSE, layers of S/Si/Ar/Ca in intermediate regions of incomplete 
burning, then, farther out, layers of O/Mg/Ne as products of explosive carbon burning. Within seconds of the explosion, these elements would be 
in a freely expanding envelope, with matter density decreasing and transparency increasing with time. It is this envelope down to the photosphere that 
we would observe, with the homologous expansion preserving the exploding body's radial structure, which is mapped onto velocity space. With the expansion, 
the photosphere recedes in velocity, revealing successively deeper layers.

In the simulations we used a spherical, delayed detonation model for an $\rm M_{Ch}$-mass WD from the 5p0z22 series of \citet{HGFS99by02}, which has been successful previously in reproducing the optical and IR light curves and spectra of a Branch-normal and several sub-luminous SNe~Ia and the statistical properties of the SNe~Ia class \citep{howell99by01,HGFS99by02,marion06a,quimby05ap07,hoeflich10,patat12}. The assumption of spherical geometry is equivalent to assuming that mixing during the deflagration phase is suppressed. The $\rm M_{Ch}$-mass WD originates from a progenitor with a main sequence mass of  5 $\rm M_\odot$ and solar metallicity. At the time of the explosion, the central density of the WD is $2\times10^9$ g cm$^{-3}$. 

Varying the amount of burning prior to the DDT produces a wide range of values for $^{56}$Ni mass and the corresponding brightness, and it shifts the characteristic chemical pattern in velocity space (see Figure 3 in \citealt{HGFS99by02}). We chose the model  5p0z22.25, which produces about 0.60 $\rm M_\odot$ of $^{56}$Ni. This model provides an absolute brightness $\rm M_B$ = -19.29 mag and brightness decline ratios $\rm \Delta$$m_{15}$(B,V) =1.02 mag and 0.61 mag in B and V, which are comparable to the observed values for SN~2014J, which has -19.19 $\pm$ 0.1 mag, $1.11 \pm 0.02$ mag, and 0.56 mag, respectively \citep{Marion14}. Our reference model explosion is marginally brighter than observed for SN~2014J, perhaps suggesting, based on the optical luminosity, that a model with slightly less $^{56}$Ni may be preferred. However, $\rm M_B$ is uncertain because of the large interstellar reddening. Moreover, $\gamma$-ray lines observed by INTEGRAL imply about 0.62 $\pm$ 0.13 $\rm M_\odot$ of $^{56}$Ni \citep{Churazov14,Isern14}.

\section{INTERPRETATION}\label{Interpretations}
\noindent \textit{\textbf{Line Identification:}}
A comparison of the synthetic and smoothed observed spectra is shown in Figure 4 assuming a distance of 3.5 Mpc to SN~2014J. The model spectra have been convolved with the spectral response function of CanariCam. The synthetic spectra are dominated by forbidden lines in emission overlaid on free-free emission, Thomson scattering, and a quasi-continuum of allowed lines formed in the central, optically thick photosphere. Due to the evolving geometrical dilution, the Thomson scattering photosphere can be identified with layers expanding at $\sim$5000 km s$^{-1}$ at 57 d and $\sim$2000 km s$^{-1}$ at 137 d. The model continuum flux density near 7.8 $\micron$ declines from about 6 to 2 mJy during this period, although the observed flux density there is zero within the measurement uncertainties.

In Table \ref{lines} we give a list of forbidden lines that contribute significantly to the emissivity at any layer. The contribution of a ``strong'' line in a specific layer to the overall emission depends on the details of the radiative transport and ionization balance and may not be significant in the observed spectra. In the model, the spectra are dominated by [Co II/III] with additional contributions by Ne, Ar, Fe and Ni. The most prominent features are the [Co III] line at 11.888 $\micron$ and [Co II] line at 10.523 $\micron$, and weaker features attributed to [Fe III] at 7.791, 8.211, 10.203 $\micron$ and [Ar III] at 8.991 $\micron$, [Co III], [Ni II] at roughly 12.7 $\micron$, and [Ni IV] at 8.405 and 11.726 $\micron$.  

In general terms, the spectra can be understood as follows. SNe~Ia are powered by radioactive decay of $^{56}$Ni $\rightarrow$ $^{56}$Co $\rightarrow$ $^{56}$Fe with half-lives of 6.1 and 77.2 days, respectively. The energy deposition that heats the gas is primarily by hard $\gamma$-rays but also, to a much lesser extent, by positrons, both of which trace the distribution of the radioactive material \citep{1994Hoeflichgamma,penney14}. By day 57 the diffusion time scale (and therefore the envelope optical depth) for low-energy photons generated in this heated gas, which scales as $t^{-2}$ due to geometrical dilution \citep{h95}, is quite short, being only about two days. Therefore, the optical depth in the forbidden lines is small, and they trace locations determined by a combination of energy deposition and critical density. During the density evolution resulting from the expansion, which scales as $t^{-3}$, the forbidden lines become stronger, and recombination time scales increase, which leads to a shift towards higher ionization.     

\noindent \textit{\textbf{Evolution of the [Co III] Line at 11.888 $\micron$:}}
Our model reproduces the evolution of the [Co III] line fairly well including the increase between 57 and 81 d and the slow evolution up to day 137. From 57 to 81 d, the photosphere recedes from layers expanding at $\sim$5900 km s$^{-1}$ to layers expanding at $\sim$3500 km s$^{-1}$ mostly due to geometrical dilution. With time, more radioactive Co gets exposed and the density drops below the critical density for collisional de-excitation. By 81 d, most of the Co is exposed in the model. Our reference model also shows a central ``hole'' in the $^{56}$Ni distribution. A shift in the initial $^{56}$Ni distribution towards the center would prolong the phase of a rising [Co III] flux, because the higher density of the $^{56}$Ni layers would delay the drop of some $^{56}$Ni below the critical density. For massive WDs, such a shift may be produced by strong mixing or initial masses less than $\sim$1.3 $\rm M_\odot$, which would avoid central electron capture. In the latter case, we would expect a narrower [Co III] line. Note that merging of very low mass WDs ($\sim$1 $\rm M_\odot$) would result in low density and thus a shorter phase of increasing [Co III].

One might expect that, after full exposure of the [Co III] region, the absolute line flux would decrease with time, and in fact we saw a decrease of about 14\% in the line flux between day 81 and day 137. However, this is a much smaller decrease than the 65\% expected from geometric dilution (which scales as $t^{-2}$). Thus, in effect, the [Co III] line flux was relatively constant between day 81 and day 137. As noted above, most of the radioactive energy is deposited by $\gamma$-rays rather than local positrons. However, there are two competing effects. 1) In our model, the absorption  probability for $\gamma$-rays decreases from 20 to 10\%, which leads to a reduction of energy input of about a factor of three. 2) The critical density for collisional de-excitation is about $(5-10)\times10^6$ cm$^{-3}$ \citep{gerardy07}, and it depends mostly on the density $\rho \propto t^{-3-n} $, with the exponent $n$ indicating the density gradient of the layers. In our reference model, $n \approx$ 1.7-2, which leads to an increase of emission in the forbidden line of about a factor of three. In fact, within our model, we expect a decreasing [Co III] flux starting at about 150 d. Note that details of the later evolution depend critically on the collisional de-excitation and the term scheme of the model atom \citep{gerardy07}. 

The line widths of [Co III] were in the range 12,600 - 13,600 km s$^{-1}$, consistent with our $^{56}${Ni} distribution (see Figure 3 in \citealt{Hoeflich:gamma2002}). The profile was rather ``stubby'' at 57 d, because the extended photosphere and high density results in a lack of emission from low-velocity material. Subsequently, it develops into a more centrally peaked profile because of the excitation of central stable Co by $\gamma$-rays, which deposit energy in the central region \citep{1994Hoeflichgamma,penney14}. Note that Co/Fe profiles in the near-IR often become ``stubby'' or ``flat topped'' again after about 300 d, because positrons dominate the energy input \citep{h04,maeda10,penney14}. However, up to about 150 days after the explosion, energy input by $\gamma$-rays dominates \citep{1994Hoeflichgamma,penney14}.

At late times, the observed [Co III] 11.888 $\micron$ line profile is well reproduced by a spherical model with respect to the line width and the shift of the line. We see no indication of large scale asymmetries in the chemical distribution of SN~2014J, which is consistent with the lack of polarization \citep{patat14}. However, at early times, the model profile lacks emission in the red wing and, consequently, the line profile is too narrow. This discrepancy may imply that the model photosphere recedes too slowly, i.e., that it has not receded quickly enough into the inner, lower-velocity regions of the outflow. We also point out that the model [Co II] feature at 10.5 $\micron$ extends too far to the blue, whereas the model [Co III] 11.888 $\micron$ profile does not.

\noindent \textit{\textbf{Comments on Other Lines:}}
In our models, the feature at $\sim$10.5 $\micron$ is dominated by [Co II], with at most a 10-15\% contribution from [S IV]. (In fact, the S does not show up in our models, but we mention it because even modest mixing can expose it.) The [Co II] feature becomes weaker with time as the ionization balance shifts towards [Co III] in the relevant region, because the recombination scales as $\rho^2$ while the energy generation and ionization scale as $\rho$ (i.e., with the radioactive decay). The same effect can be seen in the broad feature between 8.2 - 9.4 $\micron$. The red part is dominated by [Ar III]. The blue part of the broad feature, which is relatively weak in the model, may be produced by [Fe III] near 8.211 and 8.611 $\micron$ at day 57 (with those two features blended or merged together) and [Ni IV] at 8.405 $\micron$ from stable Ni starting on day 81 as the photosphere starts to penetrate that region. Note, however, that this line, if indeed it is from stable Ni, is observed to be much stronger than implied by our model.

Keep in mind that (1) the Fe traces the radioactive Ni and Co and is therefore expected to share its broad-lined velocity structure, and (2) the stable Ni to which we refer is produced in the center by electron capture and is distinct from the radioactive Ni. Note also that even a small amount of mixing of, for example, stable Ni at the Si/S interface, which is not included in the models, would boost the model line fluxes in this spectral region (see \citealt{Hoeflich:gamma2002}). 3D deflagration models predict strong mixing by plumes due to Rayleigh-Taylor instabilities during the deflagration phase. Although inconsistent with the layered structure, we may nevertheless be seeing some modest mixing by plumes. The appearance of significant stable Ni would imply that a high-density WD close to $\rm M_{Ch}$ is the likely source of SN 2014J. 

Finally we point out that, within the $\rm M_{Ch}$-mass scenarios, SN~2014J might originate from a WD with a larger central density $\sim$$(5-7) \times 10^9$ g cm$^{-3}$, which enhances electron capture and shifts the nuclear equilibrium from $^{54}$Fe, $^{57}$Co, and $^{58}$Ni in our model to $^{50}$Ti and $^{54}$Cr \citep{brachwitz00,hoeflich06}. Both Cr II and Ti II have many transitions from lower excitation levels, e.g., [Ti II] (a2G - b4P) with $\rm E_l=8997$ cm$^{-1}$, but atomic data are lacking for these (e.g., http://www.pa.uky.edu/$\sim$peter/newpage/).  

\section{SUMMARY COMMENTS}\label{DiscussionConclusions}
We present the first-ever time series of mid-IR spectra of a SN Ia. These 8 - 13 $\micron$ spectra of SN~2014J, observed at the GTC between 2 and 4 months after the explosion, demonstrate that the mid-IR can help address many of the open questions about SNe~Ia. 

1. The spectra are dominated by forbidden lines of iron group elements and Argon. The strongest of these is the [Co III] line at 11.888 $\micron$, which provides a valuable tool to monitor the time evolution of the photosphere and measure the amount of $^{56}$Ni, which drops below the critical density and, thus, probes the mass of the ejecta. SN~2014J requires a progenitor close to the $\rm M_{Ch}$ mass. Late-time line profiles of SN~2014J are rather 
symmetric and not shifted in the rest frame, consistent with the low continuum polarization of SN~2014J \citep{patat14}. Either SN~2014J has been observed along its axis 
of symmetry, or it has an overall spherical distribution of elements.

2. The mid-IR shows lines of Ar, which is a unique probe of mixing in the transition layer of products of incomplete burning and those resulting from NSE. In SN~2014J we 
do not see evidence for strong (or complete) mixing of the corresponding layers as predicted by current 3D simulations for the deflagration phase. In this regard, we also note that, as is the case for Calcium, Argon is created in layers with sufficiently high density and temperature for there to be quasi-equilibrium of the S/Si group, although it is destroyed in 
full NSE. The prominent Ca H + K lines and the near-IR triplet are optically thick in SNe~Ia even at solar abundances \citep{h95}, which makes them unusable as diagnostic tools to explore these regions. Given this lack of  suitable lines of noble gases at shorter wavelengths, the prominence of Ar in the mid-IR makes this spectral region uniquely useful for that task. 

3. Along with [Fe III] lines, we may see evidence for [Ni IV] near 8.4 $\micron$ in an, albeit highly blended, region of the spectra. In $\rm M_{Ch}$-mass models, we see the production of electron-capture elements near the center. This [Ni IV] feature is stronger than predicted by an order of magnitude. Keeping in mind the lack of atomic data for mid-IR transitions, we speculate that we may be seeing stable Ni, with central 
mixing possibly boosting that line as compared to our reference model. Alternatively, the progenitor WD may have a higher central density $\sim$$(4-6) \times 10^9$ g cm$^{-3}$, which would 
shift the NSE to Cr and Ti rather than the $^{54}$Fe, $^{57}$Co, and $^{58}$Ni produced at $\sim$$2\times 10^9$ g cm$^{-3}$. The lack of atomic data for Cr and Ti currently hinder their use as diagnostics.

4. The spectra have been interpreted using a ``classical'' delayed-detonation model for a Branch-normal SNe~Ia with values of parameters, including 0.60 $\rm M_\odot $ of $^{56}$Ni, that are consistent with the  absolute brightness and the brightness decline relation observed for SN~2014J \citep{Marion14}. The inferred mass of $^{56}$Ni is in agreement with values derived from  $\gamma$-ray fluxes observed for SN2014J \citep{Isern14,Churazov14}, thus providing a consistent and completely independent check on the value of this fundamental parameter. Overall, the models reproduce the observations of SN~2014J, including their temporal evolution. 

Mid-IR observations provide a new and exciting window through which we may understand SNe~Ia. We are at the beginning of exploring this wavelength region, and it is providing some answers but also new questions. Our first spectra of SN 2014J were taken nearly two months after the explosion, but catching other SN Ia in the mid-IR earlier than this would be very informative. For example, we expect the $^{56}$Ni layers to emerge near maximum light, with the Co and Ar lines starting to appear at about one month after the explosion \citep{h95}. Among other constraints, the onset of these and other lines would provide critical insight into mixing processes. We also note that spectra spanning more than a few months after the explosion are needed to study the central region and the transition from the Co-line-dominated to the Fe-line-dominated 
regimes, and we will attempt to obtain those for SN~2014J over the next year. Including our observations, SN~2014J will be one of the best observed SNe~Ia from the UV, optical, near-IR and mid-IR to the radio-range. Finally, we emphasize that the diversity of SNe~Ia has started to become apparent during the last few years. SNe~Ia within about 10 Mpc, which are expected to appear with a frequency of about one per year (http://www.rochesterastronomy.org/sn2013/snredshift.html), are now within reach of a powerful mid-IR instrument like CanariCam that can probe that diversity.

\acknowledgments
{
We would like to thank many colleagues and collaborators for their support. CMT wishes to acknowledge particularly useful comments by David Collins. The work presented in this paper has been supported in part by NSF awards AST-0708855 and AST-1008962 to PAH and NSF awards AST-0903672 and AST-0908624 to CMT. CMW acknowledges support from ARC Future Fellowship FT100100495. We would also like to express our thanks to Peter van Hoof for creating the Atomic Line List V2.05B18 at http://www.pa.uky.edu/ peter/newpage/ and to the GTC staff for their outstanding support of the commissioning and science operations of CanariCam. This research is based on observations using CanariCam at the Gran Telescopio Canarias, a partnership of Spain, Mexico, and the University of Florida, and located at the Spanish Observatorio del Roque de los Muchachos of the Instituto de Astrof\'{i}sica de Canarias, on the island of La Palma.}

\clearpage

\begin{figure}   
\includegraphics[width=0.90\textwidth]{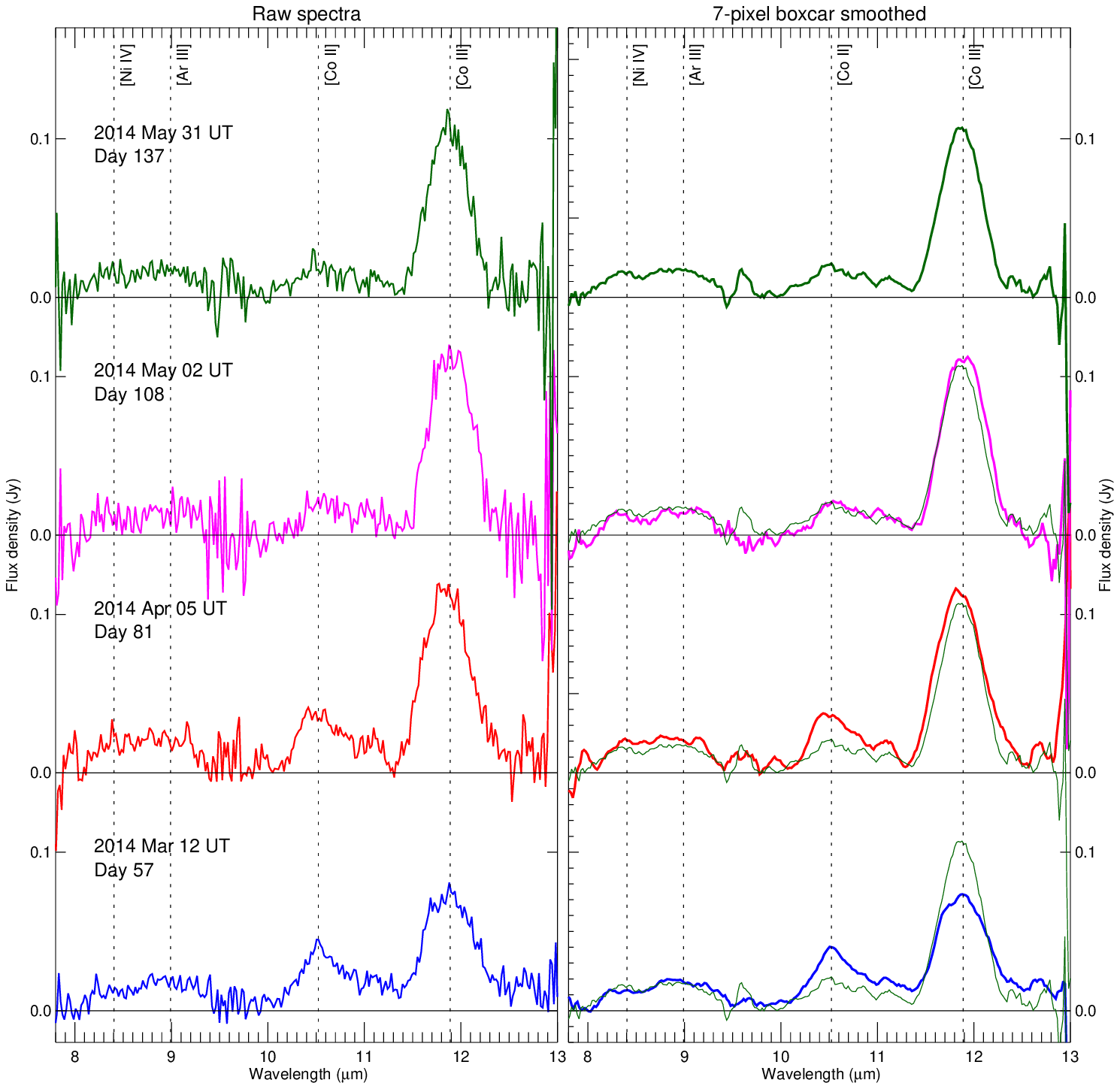}
\caption{Spectral evolution of SN2014J in the mid-IR at four epochs between 2014 March 12 and May 31 as observed with CanariCam, with original data displayed in left panels and 7-pixel (0.14 $\micron$) boxcar-smoothed data in right panels. The smoothed spectrum obtained on day 137 (thin lines in all right panels) is plotted for reference. Some candidate spectral features are indicated. The spectral resolution is $\sim$0.15 $\micron$ corresponding to a Doppler shift of $\sim$4090 km s$^{-1}$ at 11 $\micron$, which is several times smaller than the observed line widths. (A color version of this figure is available in the online journal.)}
\label{fig:spec_obs}
\end{figure} 

\begin{figure}
\begin{center}
\includegraphics[width=0.8\textwidth]{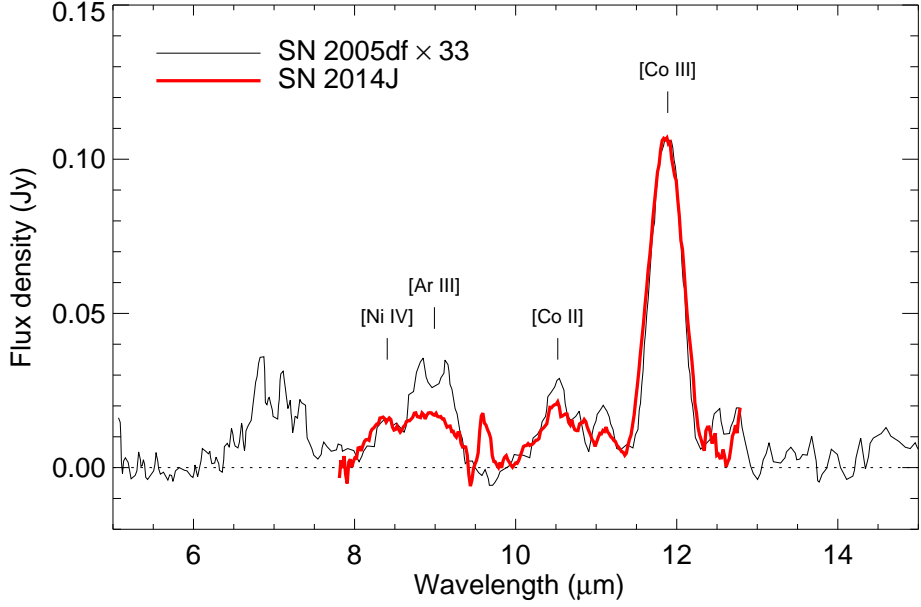}
\caption{Comparison of spectrum of SN 2005df obtained with \it Spitzer \rm on day 135 by \citet{gerardy07} to that of SN 2014J obtained by us on day 137. The flux of SN 2005df has been scaled by a factor of 33. (A color version of this figure is available in the online journal.)}
\end{center}
\label{fig:compare}
\end{figure} 

\begin{figure}
\begin{center}$
\begin{array}{cc}
\includegraphics[width=0.45\textwidth]{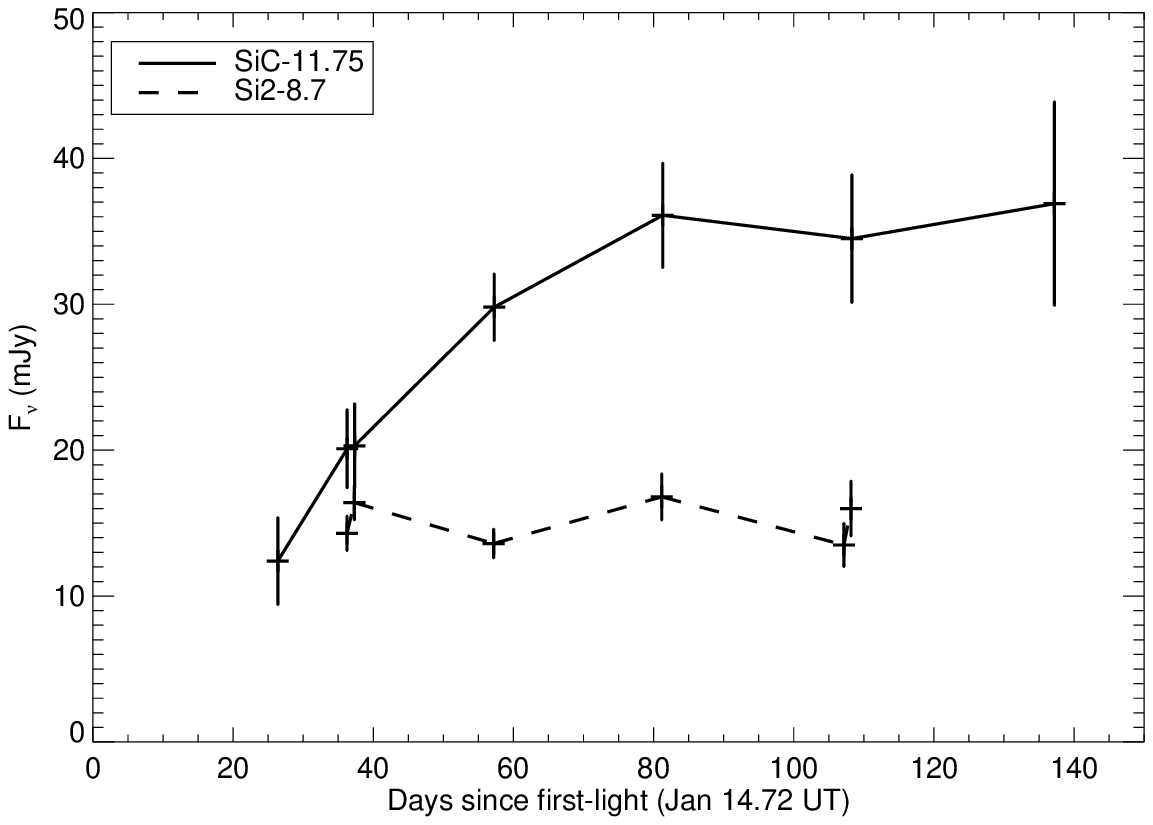} &
\includegraphics[width=0.45\textwidth]{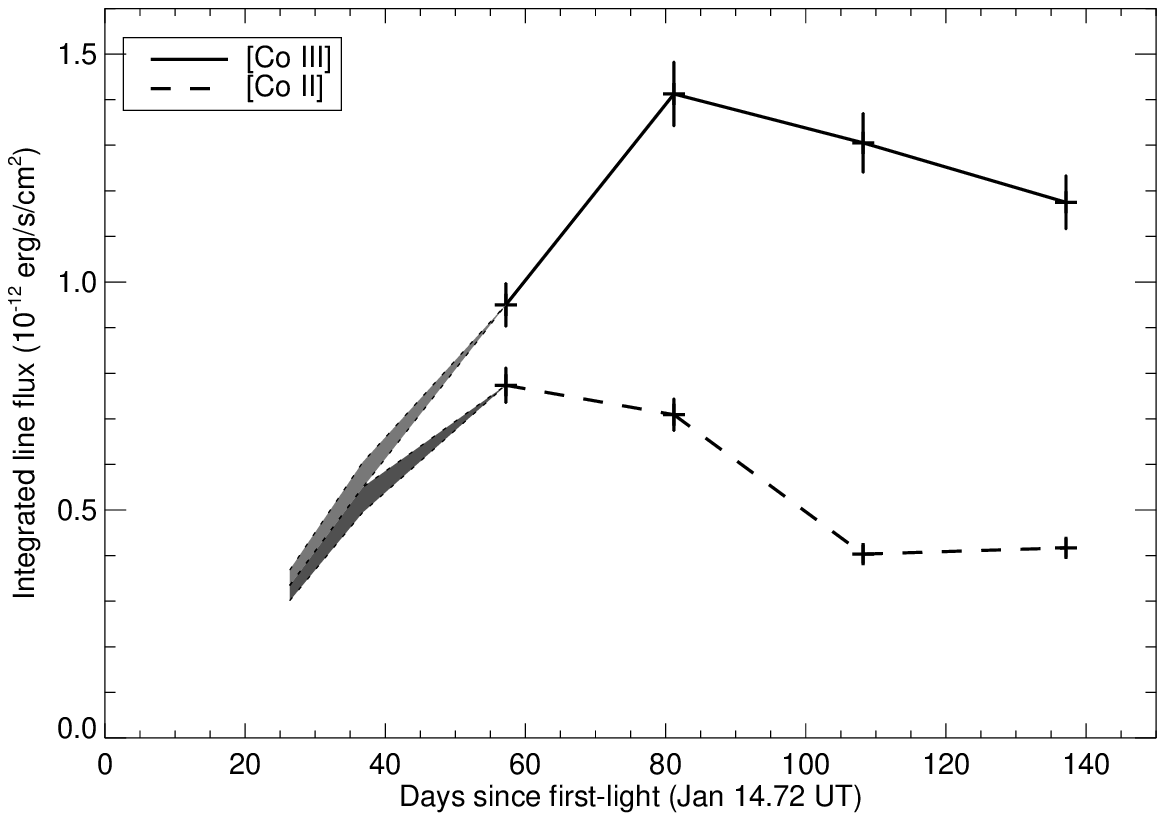}
\end{array}$
\caption{Temporal variations of fluxes in broad band filters (left panel) and corresponding fluxes in the [Co II] 10.523 $\micron$ and [Co III] 11.888 $\micron$ lines (right panel). Line fluxes for days 26 and 36/37 are estimated from the filter photometry of the spectral region spanning those lines (see text). Line fluxes for days 81 and later are derived directly from the spectra.}
\end{center}
\label{fig:broad}
\end{figure}

\begin{figure}
\begin{center}
\includegraphics[width=0.8\textwidth]{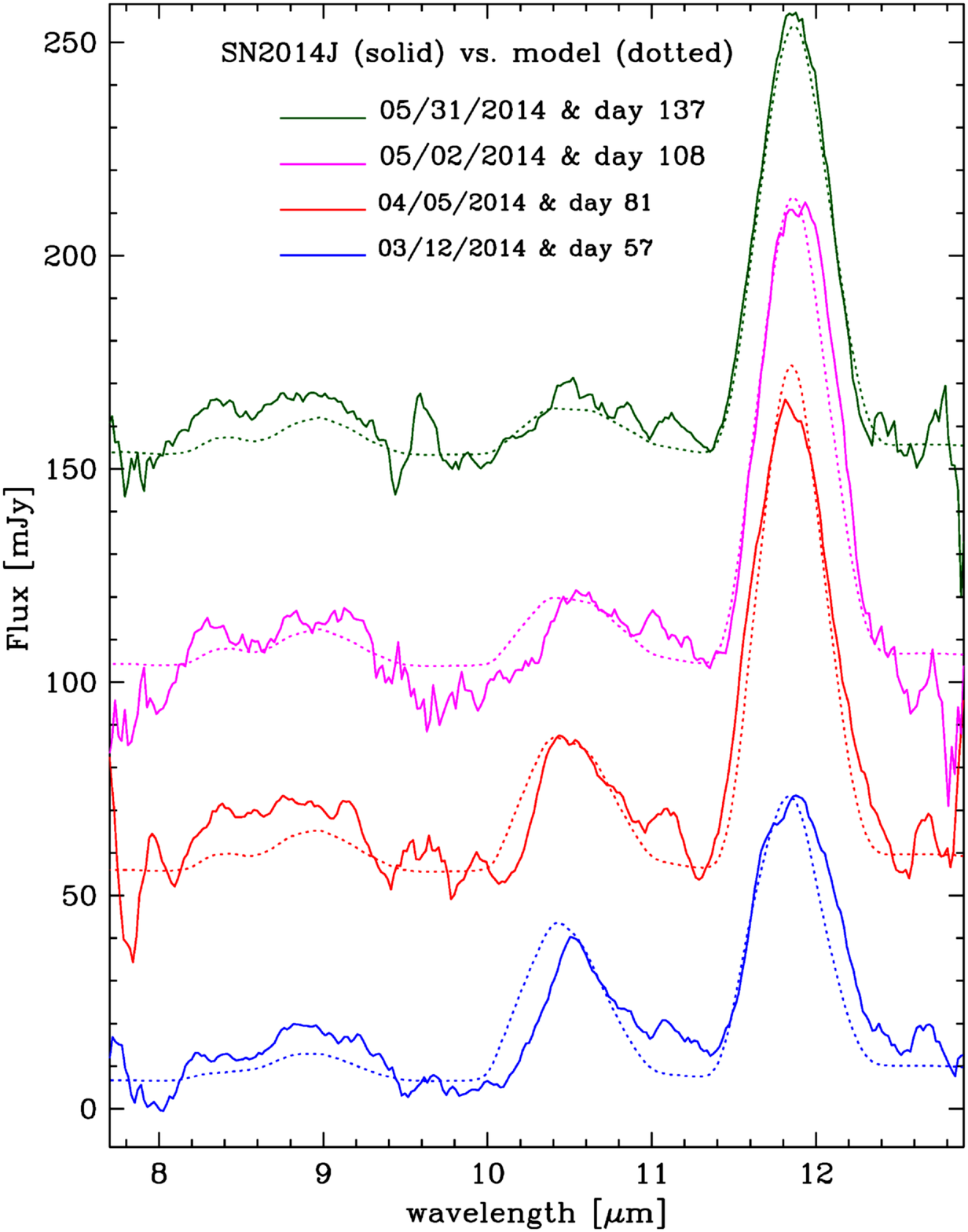} 
\caption{Comparison of the observed mid-IR spectra (7-pixel smoothed, as in Figure 1) of SN 2014J to synthetic spectra (dotted lines) produced using the reference model. (A color version of this figure is available in the online journal.)} 
\end{center}
\label{fig:synspec}
\end{figure}

\clearpage

\begin{deluxetable}{lcccc}
\tabletypesize{\scriptsize}
\tablecaption{Observing Log and Flux Densities\label{tab:log}}
\tablewidth{0pt}
\tablehead{
\colhead{UT Date\tablenotemark{a}} & \colhead{Age\tablenotemark{b}} & \colhead{Filter/Grating} & \colhead{Integration Time} & \colhead{$F_{\nu}$} \\
\colhead{ }                                               & \colhead{(days)}                             & \colhead{ }                        & \colhead{(s)}                           & \colhead{(mJy)}
}
\startdata
\multicolumn{5}{c}{Imaging} \\
\hline
Feb 10.12 & 26.40  & SiC-11.75 & 2 $\times$ 200 & 12.4 $\pm$ 3.0 \\
Feb 10.13 & 26.41  & Si4-10.3 & 2 $\times$ 200 & 7.2 $\pm$ 2.9 \\
Feb 20.00 & 36.28  & Si2-8.7 & 4 $\times$ 200 & 14.3 $\pm$ 1.2 \\
Feb 20.03 & 36.31  & SiC-11.75 & 4 $\times$ 200 & 20.1 $\pm$ 2.7 \\
Feb 21.05 & 37.33  & Si2-8.7 & 4 $\times$ 200 & 16.4 $\pm$ 1.2 \\
Feb 21.08 & 37.36  & SiC-11.75 & 4 $\times$ 200 & 20.3 $\pm$ 2.9 \\
Mar 12.93 & 57.21  & Si2-8.7 & 550 &  13.6 $\pm$ 1.0 \\
Mar 13.01 & 57.29  & SiC-11.75 & 550 & 29.8 $\pm$ 2.3\\
Apr 05.92 & 81.20  & Si2-8.7 & 410 & 16.8 $\pm$ 1.6\\
Apr 06.05 & 81.33  & SiC-11.75 & 410 & 36.1 $\pm$ 3.6 \\
May 01.92 & 107.20 & Si2-8.7 & 350 & 13.5 $\pm$ 1.5 \\
May 02.93& 108.21 & Si2-8.7 & 350 & 16.0 $\pm$ 1.9 \\
May 03.05& 108.33 & SiC-11.75 & 350 & 34.5 $\pm$ 4.4 \\
\hline
\multicolumn{5}{c}{Long-slit spectroscopy} \\
\hline
Mar 12.95 & 57.23  & LowRes10 & 4 $\times$ 650 & - \\
Apr 05.95 & 81.23  & LowRes10 & 4 $\times$ 530 & - \\
May 02.95 & 108.23 & LowRes10 & 4 $\times$ 470 & - \\
May 31.89 & 137.17 & LowRes10 & 4 $\times$ 470 & - \\
\enddata
\tablenotetext{a}{UT date at the start of integration.}
\tablenotetext{b}{In days from the explosion assumed to have been on Jan 14.72 UT \citep{zheng2014}.}
\end{deluxetable}

\begin{deluxetable}{ccccccccc}
\tablewidth{0pt}
\tabletypesize{\scriptsize}
\tablenum{2}
\tablecaption{Strong Mid-IR Forbidden Lines in Reference Model\label{lines}}
\tablehead{
\colhead{$\lambda$ [$\micron$]} &
\colhead{Str.}&
\colhead{Ion}&
\colhead{Configuration}&
\colhead{Term}&
\colhead{$J_{l,u}$}&
\colhead{$A_{ul}$}&
\colhead{$g_u A_{ul}$}&
\colhead{$E_{l,u}$ [cm$^{-1}$]}}
\startdata
7.791&$++$&[Fe~III]&3d6-3d6&3P4-3P4&2-1&4.70E-02&1.41E-01&19404-20687\\
8.211&$+$&[Fe~III]&3d6-3d6&3H-3F4&4-3&1.50E-08&1.05E-07&20481-21699\\
8.405&$++$&[Ni~IV]&3d7-3d7&4F-4F&9/2-7/2&5.70E-02&4.56E-01&0-1189 \\
8.611&$+$&[Fe~III]&3d6-3d6&3H-3F4&5-4&9.50E-09&8.55E-08&20300-21462\\
8.991&$++$&[Ar~III]&3s2.3p4-3s2.3p4&3P-3P&2-1&3.10E-02&9.30E-02&0-1112\\
9.723&&[Ni~IV]&3d7-3d7&2H-2H&11/2-9/2&1.60E-02&1.60E-01&26649-27677\\
9.975&&[Ni~IV]&3d7-3d7&2P-2P&3/2-1/2&1.70E-02&3.40E-02&23648-24651\\
10.080&&[Ni~II]&3d8.(1D).4s-3d8.(3P).4s&2D-4P&3/2-3/2&1.20E-02&4.80E-02&23796-24788\\
10.088&&[Fe~III]&3d6-3d6&1S4-1D4&0-2&8.10E-10&4.05E-09&34812-35803\\
10.203&$+$&[Fe~III]&3d6-3d6&3H-3F4&4-4&1.60E-03&1.44E-02&20482-21462\\
10.523&$++$&[Co~II]&3d8-3d8&a3F-a3F&4-3&2.23E-02&1.56E-01&0-950\\
10.682&$+$&[Ni~II]&3d8.(3F).4s-3d8.(3F).4s&4F-4F&9/2-7/2&2.71E-02&2.17E-01&8394-9330\\
11.002&&[Ni~III]&3d8-3d8&3F-3F&3-2&2.70E-02&1.35E-01&1371-2270\\
11.167&&[Co~II]&3d7.(4F).4s-3d7.(4F).4s&b3F-b3F&4-3&1.87E-02&1.31E-01&9813-10708\\
11.726&$++$&[Ni~IV]&3d7-3d7&4F-4F&7/2-5/2&3.60E-02&2.16E-01&1190-2043\\
11.888&$+++$&[Co~III]&3d7-3d7&a4F-a4F&9/2-7/2&2.00E-02&1.60E-01&0-841\\
11.978&&[Fe~III]&3d6-3d6&3P2-3P2&1-2&7.90E-03&3.95E-02&49576.8-50412\\
12.641&&[Fe~II]&3d7-3d7&a2D2-a2D2&5/2-3/2&8.01E-03&3.20E-02&20517-21308\\
12.681&$+$&[Co~III]&3d7-3d7&a2G-a2G&9/2-7/2&7.20E-03&5.76E-02&16978-17766\\
12.729&$+$&[Ni~II]&3d8.(3F).4s-3d8.(3F).4s&4F-4F&7/2-5/2&2.76E-02&1.66E-01&9330-10116\\
12.814&$++$&[Ne~II]&2s2.2p5-2s2.2p5&2Po-2Po&3/2-1/2&8.32E-03&1.66E-02&0-780\\
\enddata
\tablecomments{We give the wavelength $\lambda$ ($\micron$), a qualitative line strength in specific layers of the model (+,++,+++), the ion, configuration, term, momentum $J_{l,u}$ of the lower and upper level, Einstein's $A_{ul}$ and $g_k A_{ul}$, and the lower and upper energy level $E_{l,u}$ (cm$^{-1}$). Note that some transitions are strong in some layers but not globally, resulting in overall weak features in the spectra.}
\end{deluxetable}

\end{document}